# Chemical Heredity as Group Selection at the Molecular Level

Omer Markovitch[1,2], Olaf Witkowski[3] and Nathaniel Virgo[3]

((1) Center for Systems Chemistry, Stratingh Institute, University of Groningen, Groningen, The Netherlands, Email: omermar@gmail.com ; (2) Blue Marble Space Institute of Science, Seattle, Washington, United States of America ; (3) Earth-Life Science Institute, Tokyo Institute of Technology, Tokyo, Japan, Emails: {olaf.witkowski@gmail.com, nathanielvirgo@gmail.com})

**Abstract**

Many examples of cooperation exist in biology. In chemical systems however, which can sometimes be quite complex, we do not appear to observe intricate cooperative interactions. A key question for the origin of life, is then how can molecular cooperation first arise in an abiotic system prior to the emergence of biological replication. We postulate that selection at the molecular level is a driving force behind the complexification of chemical systems, particularly during the origins of life.

In the theory of multilevel selection the two selective forces are: within-group and between-group, where the former tends to favor "selfish" replication of individuals and the latter favor cooperation between individuals enhancing the replication of the group as a whole. These forces can be quantified using the Price equation, which is a standard tool used in evolutionary biology to quantify evolutionary change. Our central claim is that replication and heredity in chemical systems are subject to selection, and quantifiable using the multilevel Price equation.

We demonstrate this using the Graded Autocatalysis Replication Domain computer model, describing simple protocell composed out of molecules and its replication, which respectively analogue to the group and the individuals. In contrast to previous treatments of this model, we treat the lipid molecules themselves as replicating individuals and the protocells they form as groups of individuals.

Our goal is to demonstrate how evolutionary biology tools and concepts can be applied in chemistry and we suggest that molecular cooperation may arise as a result of group selection. Further, the biological relation of parent-progeny is proposed to be analogue to the reactant-product relation in chemistry, thus allowing for tools from evolutionary biology to be applied to chemistry and would deepen the connection between chemistry and biology.

## 1. Introduction

A biological cell consists of a complex interlined and interdependent network of many processes, including the metabolic construction of molecular building blocks; the assembly of proteins; the replication of nucleic acids; and the production of and transport across the cell's membrane. In chemistry, however, even though complex chemical systems are known (Skene and Lehn 2004, Mattia and Otto 2015) and some of which have been found to give rise to the central phenomena of replication (von Kiedrowski, Otto et al. 2010, Ashkenasy, Hermans et al. 2017), they do not approach the complexity of biology.

Biological systems are highly cooperative, with the example of the cell with its many molecular constituents, including proteins, RNA and DNA. In chemical systems, molecular cooperation may be more difficult to study and understand. We study the emergence of cooperation in chemical systems, postulating that molecular selection can drive complexification in such system. To do this, we reach out to biological theories that have built up a large amount of literature on this topic throughout the years, to understand the emergence of cooperation in such systems. The combination of biological and chemical theories for cooperation may then shed light on new ways to approach these phenomena at different scales.

A common way to quantify group selection in biology is by applying the Price equation (Frank 2012), which is presented in section 3. After that, we will discuss on how to analyze chemical systems with the Price equation (section 4), aiming to focus on chemical cooperation. Finally, we will argue that a well-studied assembly-based model, GARD (Segre, Ben-Eli et al. 2000, Segre and Lancet 2000), may be cast in terms of multilevel selection, with the two levels being individual lipid molecules and assemblies thereof (sections 5 and 6).

## 2. Cooperation in biology

Cooperation in biology is a behavior that provides a benefit to a recipient at a cost for the cooperator. This behavior is selected for because of its beneficial effect on the recipient (Hamilton 1964, Dawkins 1979). A general umbrella of theories of the emergence of cooperation in biological systems fall under group selection theory, recently revisited under the denomination of multilevel selection theory (Okasha 2006, Traulsen and Nowak 2006). Cooperative effects are not explained by traditional evolutionary theory, and alternatives such as "cultural group selection" or "gene-culture co-evolution" have been introduced to explain them in the case of human culture (Bowles, Choi et al. 2003, Fehr and Fischbacher 2003).

## 2.1. Group selection

Group selection is a mechanism in evolutionary biology in which natural selection acts more dominantly at the level of the groups rather the individuals (Smith 1964, Dawkins 1979, Grafen 1984, West, Griffin et al. 2007). This group selection can be due to interacting individuals within the group which provides additional fitness beyond that of the individuals. The effect may be significant as groups can compete in an analogue fashion to how individuals compete, creating a system where selection can act on multiple layers.

Group selection theory (Wilson and Sober 1994) — also called new group selection theory — has been showed to be mathematically equivalent to kin selection (Frank 1986, Bourke and Franks 1995, Wenseleers, Helanterä et al. 2004), a term coined by Maynard Smith (Smith 1964), and has undergone some debate (West, Griffin et al. 2007). Kin selection is the evolutionary process by which some traits are favored because of their beneficial effects on the fitness of the relatives (West, Griffin et al. 2007), also called inclusive fitness. Kin selection theory states that the coefficient of genetic relatedness, often taken as the probability that a gene picked randomly from each at the same locus is identical by descent, must exceed the cost-to-benefit ratio of the altruistic act: $r*B>C$ (where $r$ is genetic relatedness, $B$ is the additional reproductive benefit gained by the recipient of the altruistic act, and $C$ is the reproductive cost to the individual performing the act), which is known as Hamilton's rule (Hamilton 1964). The theories tell us that increasing the group benefits and reducing the individual cost favors cooperation, and yield identical results for similar situations (Okasha 2006).

Multilevel selection theory revisits group selection by considering groups as functional targets for selection, with the evolution acting at the level of groups of individuals within a species (Figure 1). The balance between group-selection and individual-selection can be evaluated in specific cases. For example, the theory suggests a generalization of relatedness that does not require Hamilton's original assumption of direct genealogical relatedness (Wilson and Wilson 2007).

A common way to quantify group selection in biology is by applying the Price equation, which is presented in section 3.

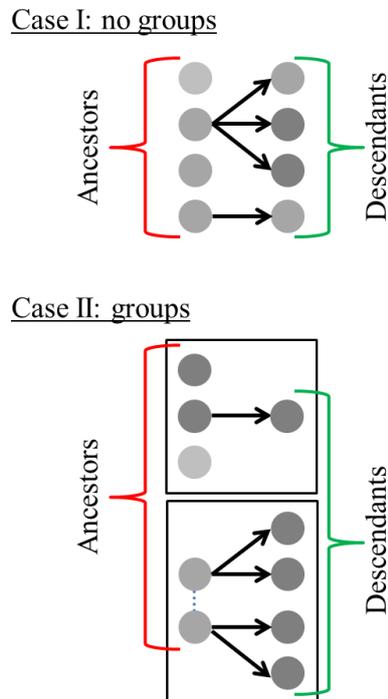

**Figure 1: Group selection heuristic.** Circles represent individuals, with the value of the trait (z, darker shading means higher value) proportional to the color scale. Case I : When no groups exists, individuals compete with each other. Case II : When groups exist, selection can favor a particular group because cooperation between parents (represented by the dotted line) leads to more descendants in this group.

### 3. Price equation

Evolutionary biology studies how individuals and populations change. Individuals multiply (replicate) via fission and budding. Knowing which parent gave rise to which progeny allows for understanding how the fitness of an individual correlates with its own progeny's fitness. Ultimately, this leads to a description of how evolution takes place, as the offspring resulting from the multiplication are often imperfect copies of their parents.

The Price equation breaks the change in a trait (z) between two *related* populations (ancestor and descendant) into two parts: selection and transmission (Price 1970). This is often written as $\bar{w}\Delta z = cov(w, z) + E(wz')$. Here we will write it explicitly as:

$$\frac{\sum_i z_i^D w_i}{\sum_i w_i} - \langle z_i^A \rangle = \frac{1}{\langle w_i \rangle}\left(cov(w_i, z_i^A) + \langle w_i(z_i^D - z_i^A)\rangle\right)$$

Equation 1

Where for ancestor i, $w_i$ is its number of descendants, $z_i^A$ its trait value and $z_i^D$ is the average trait value among its descendants. $\langle \circ \rangle$ denotes an average, and cov denotes a covariance ($cov(x_i, y_i) = \langle x_i \cdot y_i \rangle - \langle x_i \rangle \cdot \langle y_i \rangle$). The left-hand side of Equation 1 gives the change in the average trait value between the two populations, with the first term being equal to the average trait value across all descendants. The covariance term is taken to represent selection.

When the population is divided into groups (Figure 1), Equation 1 can be written (Price 1972, Kentzoglanakis, Brown et al. 2016) to take the existence of these groups into account:

$$\langle w_{ij} \rangle \langle \Delta z \rangle = cov_i(\langle w_i \rangle, \langle z_i^A \rangle) + \langle cov_j(w_{ij}, z_{ij}^A) \rangle_i + \langle w_{ij}(z_{ij}^D - z_{ij}^A) \rangle_{ij}$$

Equation 2

Where now i is the index of a group and j the index of individual within a group. The first two terms on the right side of Equation 2 respectively represent between-groups and within-group selection, while the third one represents transmission during replication. For an ancestor individual j at group i, $w_{ij}$ is the number of descendants it gave rise to, $z_{ij}^A$ is its own z value and $z_{ij}^D$ is the average trait value among its descendants. $\langle \Delta z \rangle$ is the total observed change in z between the ancestors to descendants populations, and $\langle w_i \rangle$ and $\langle z_i^A \rangle$ are the average w and z values of all members of ancestor group i. The details of each term are given in the Appendix.

As emphasized by Frank (Frank 2012), the Price equation embodies very few assumptions about the actual biology. To use this equation to quantify selection we need only the following four things: an ancestor population, a descendant population, the trait value of each member in each population, and a many-to-one mapping from descendants to ancestors.

(Kerr and Godfrey-Smith 2009) have extended this many-to-one mapping to more general relations thus accounting for more processes, such as migration or multiple parents per progeny.

As Price himself wrote (Price 1995), the notion of *selection* is distinct from notions of reproduction and heredity, and selection can be applied to collections of inanimate objects as well as to replicating/reproducing organisms. This enables asking what kind of role selection plays in inanimate chemical systems. (Guttenberg, Virgo et al. 2017) proposed a model in which selection at the molecular level is frequency-dependent, which provides sufficient nonlinearity for changes to accumulate over time even though the molecules themselves do not reproduce. Next, we discuss how can the Price equation be applied in chemistry.

## 4. **Parent-progeny vs. reactant-product**

As discussed above, the Price equation is a definition of selection, and it is applicable in any context where it makes sense to talk about ancestors, descendants and trait values, or concepts that can be mapped to these. We are arguing in the present manuscript that it is possible to establish an analogy between the parent-progeny relationship in biology, and the reactant-product relationship in chemistry (Table 1). In chemistry, it is less clear how to apply concepts of biological evolution, mainly due to two factors, which we will detail further below: the time scales of chemical reactions and the notion of "generation".

- Chemical reactions' rates can range over many orders of magnitudes and reactions sometimes occur in a less synchronous manner. In theory, as long as it is possible to track or trace which copies of which reactants gave rise to which copies of products the distribution of reactions rates should be of little consequences. However, a heterogeneous system with multiple reactions that span over seconds, hours and days in practice is likely to pose challenges that may require sophisticated laboratory instrumentation and analyses. Enzymes, however, are known to create synchronization, in the sense that while un-catalyzed reaction rates can span over 15 orders of magnitude, the enzymatically catalyzed rates only span over 5 orders of magnitude (Wolfenden and Snider 2001). Presently, the use of computer simulations allows tracking a system with plethora of reactions and chemicals.

- While the notion of a "generation" is relatively well defined in biology, it is less so in chemistry. In biology, taking the example of the cell, there is often a clear distinction between the processes of growth and division, where the latter directly leads to the creation of progeny. In a chemical system, where multiple reactions are constantly occurring and at different time-scales, the notion of a "chemical generation" seems more difficult to define. In order to define chemical generation time we seek analogy from the biological generation time, which is commonly defined as the average time it takes for a biological population to double in numbers (Coale 2015). We thus suggest a working definition for chemical generation time, which is the time it takes for a system to double in *mass*. A caveat in this definition is that it does not always necessitates the production of a progeny.

We also note that one has to pay attention to chemical reversibility. Chemical reversibility is the notion that if a forward reaction R→P is possible, then the backward reaction R←P is also possible. The case where products give rise to reactants, but not to exactly the same molecular copies (i.e. $R_x \rightarrow P_y \rightarrow R_z$, where indices x, y and y symbolize identity), is analogue to a progeny give rise to an individual that is resembling or even identical to the original parent, but is not *the* actual parent, and has been discussed by (Kerr and Godfrey-Smith 2009). We note a special case where the products give rise to exactly the same molecular copies of the parents (i.e. $R_x \rightarrow P_y \rightarrow R_x$), where to the best of our knowledge does not occur in biology.

With the above-mentioned issues in mind, we suggest two chemical systems in which we believe it is possible to apply analysis from evolutionary biology and identify reactant-product as equivalent of parent-progeny. More examples could be exist. Here we will focus on the second example of self-assembly as a mean to demonstrate how evolutionary biology concepts could be applied in chemistry.

- Polymerization is a process that can be generally written in a simplified manner as elongation $A_n + A_m \rightarrow A_{n+m}$ and cleavage $A_x \rightarrow A_k + A_{x-k}$ (note that the second reaction is not necessarily the reverse of the first reaction), where $A_i$ is a polymer with i individual units (length). We propose that in such a system the molecules $A_n$ and $A_m$ are the ancestors of $A_{n+m}$ and the molecule $A_x$ is the ancestor to both $A_k$ and $A_{x-k}$. In applying the Price equation in this system one can consider that only single monomer additions can occur (that is, n or m = 1). It is also possible to extend the Price equation to account for multiple parents (Kerr and Godfrey-Smith 2009) (which is beyond the scope of the present manuscript), thus avoiding the need to consider n or m = 1. It is also possible to extend this example to hetero polymers.
- Self-assembly is the accretion of components into larger structures (assemblies). Perhaps the most common self-assembly is of amphiphiles spontaneously forming vesicles in aqueous solution. Vesicle-forming amphiphiles have been shown to exist in meteorites and are relevant for the origins of life (Luisi, Walde et al. 1999, Chen and Walde 2010), and a highly heterogeneous vesicle/assembly system have been proposed as an alternative to "alphabet" inheritance of bio-polymers such as DNA, RNA, etc (Segre and Lancet 2000). Furthermore, vesicles have been shown to be subjected to fission and alike processes, which makes it somewhat conceptually easier to separately think of the growth and split processes. The present work analyses a vesicle system from the point of view of the Price equation.

| Biology | Chemistry | Self-Assembly Model |
|---------|-----------|---------------------|
| Parent | Reactant | Recruiting particle |
| Offspring | Product | Recruited particle |
| Reproduction event | Reaction | Recruitment of a new particle |
| Organism | Molecule | Particle |
| Group | Assembly / Complex / Coacervate / etc | Vesicle / Micelle |

**Table 1: Entities in biology and chemistry.**

## 5. GARD model

With the above discussion in mind, we employ a computer model in order to demonstrate the study of group selection in chemistry. The GARD (Graded Autocatalysis Replication Domain) stochastic model describes the growth and fission of molecular assemblies, consisting of a large repertoire (alphabet) of simple molecules (size $N_G$) (Segre, Ben-Eli et al. 2000, Segre and Lancet 2000, Segre, Ben-Eli et al. 2001). Importantly, these assemblies store information in the form of non-random molecular compositions – compositional information (i.e. the specific ratio of different molecule-types that make up the assembly) – and pass it to progeny via homeostatic growth accompanied by fission (Gross, Fouxon et al. 2014, Markovitch and Krasnogor 2018). It is this fission that generates progeny. Molecules from the environment join an assembly and once the number of molecules in an assembly reached a predefined maximal size ($N_{max}$) a random fission event takes place producing two daughter assemblies of the same size ($N_{max}/2$). The rates at which different molecules are joining an assembly are biased by a matrix (termed β) which holds the assumed interaction between different molecule-types. This matrix can be represented as a Kauffman-like network whose $N_G$ nodes are molecule-types and $N_G^2$ edges are the strengths of the interactions. The content of each assembly (i.e. its composition) is written as a vector holding the counts of each molecule-type, and the degree of compositional similarity between two assemblies is often calculated as the dot-product between their vectors.

In the very basic model simulation, after a split event one of the daughters is discarded (picked at random) thus following the fate of a single assembly at any given time. In this case, an assembly which has similar content (high compositional similarity) to its parent and followed-progeny is considered as faithfully-replicating and termed composome (for compositional genome (Segre, Ben-Eli et al. 2000)). All the composomes of a long simulation are recorded and clustered into composome-types (compotypes), as detailed in (Shenhav, Oz et al. 2007). A compotype is represented by a vector constituting the center of mass of all its composomes. Compotypes are regarded as GARD's species and have been shown to respond to selection (Markovitch and Lancet 2012) and obey the quasispecies formalism (Gross, Fouxon et al. 2014). For a discussion of a dispute regarding selection in GARD see (Markovitch and Krasnogor 2018).

When performing population-dynamics of GARD assemblies, a constant size environment (chemical reactor) is assumed which holds a fixed number of assemblies ($L_{pop}$) at any given moment. Each of the assemblies in the population is simultaneously growing and when an assembly undergoes split, another assembly (randomly picked) is removed from the system, as detailed in (Markovitch and Lancet 2014).

GARD's equations (Segre, Ben-Eli et al. 2000) take a mean-field approximation by assuming of a fast pre-equilibrium for the dynamic collisions between the joining/leaving molecule and all other molecules of the assembly (Segré, Pilpel et al. 1998). For the sake of the present work, GARD dynamics will be considered only in the forward (joining) manner and be written as:

$$\frac{dn_i}{dt} = k_f \rho_i \beta_{ij}$$

Equation 3

Where Equation 3 models the current rate at which a copy of molecule-type i (i=1..$N_G$) is joining an assembly. $n_i$ is the number of copies of molecule-type i currently in the assembly, $k_f$ is basal joining rate constant, $\rho_i$ is environmental concentration (often taken to be buffered and uniform amongst all i) and $\beta_{ij}$ is the rate-bias (catalysis) exerted by assembly dwelling molecule-type j on type i.

Typically, $\beta_{ij}$ values are randomly drawn from a log-normal distribution (Segre, Shenhav et al. 2001) and once drawn are fixed during a simulation. Because $\beta_{ij}$ are randomly drawn, a rigorous study of GARD requires sampling many different β instances, each of which represent different potential environmental prebiotic chemistries.

The reason behind the specific form of Equation 3 is that it allows us to unequivocally track which individual molecule within an assembly drew which individual molecule from outside the assembly (Figure 2). This in turn allows us to establish an equivalent to parent-progeny in this case as j–i respectively (see Table 1), referring to *specific copies* of j and i. This also makes the model resemble even further to autocatalytic sets (Hordijk and Steel 2015).

When performing computer simulations, Equation 3 in population mode is simulated using parameter values identical to those employed in previous studies (Markovitch and Lancet 2012, Markovitch and Lancet 2014, Guttenberg, Laneuville et al. 2015): $N_{max}=10^2$, $N_G=10^2$, $k_f=10^{-2}$ and $\rho_i=10^{-2}$. We provide the full history of the trajectories, as well as the underlying $\beta$ and the compotype ((Markovitch, Witkowski et al. 2018) https://zenodo.org/record/1172467).

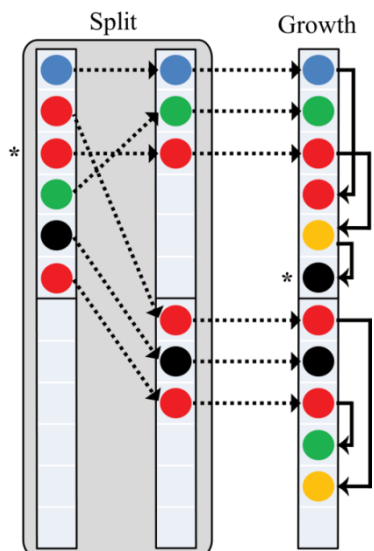

**Figure 2: Tracking the parent-progeny relationship in GARD.** In this heuristic assembly maximal size ($N_{max}$) = 6. Arrows represent how the identity of a parent is recorded: when an assembly splits, its molecules are transferred to its offspring and the source of each molecule is recorded (arrows with dotted lines); During growth, the molecule that drew each molecule into the assembly is recorded (arrows with full lines). For example, the original parent of the black molecule is the red molecule (both marked with asterisk).

## 6. Results

This section will show examples of GARD's population dynamics (under non-reversible joining, Equation 3) and how the multilevel Price equation (Equation 2) was applied to the data. The results presented below are not meant to generalize on GARD since only a single, random GARD chemistry ($\beta$) was studied. Rather, we aim to demonstrate how group selection can be studied in a chemical system via the Price equation.

### 6.1. Population dynamics

All simulations performed under the same $\beta$ and with identical GARD parameters, except for different population size, which correspond to different numbers of assemblies in the population. Under this $\beta$ two composome-types (compotypes) emerge, and for the sake of the present demonstration only one them is considered (Figure 3). In each population at each point in time, the count of the compotype is the number of assemblies with high content similarity to it. Figure 4 shows snapshots of the population dynamics. The average frequency (Figure 5) increases from about 0.7 to 0.85 when the population size increases from 3 to 10, and for larger population sizes the average frequency is at a plateau value of about 0.85. The reason for showing results for population sizes 3 and above, is that for smaller population sizes the particular population-simulation-mode used here (both daughters of a splitting parent are kept) introduces a bias. For a population size of 2, the bias is that every time one assembly splits the other assembly is removed in order to keep the population size constant, so that the lineage of the population after the split comes only from the splitting assembly. For population size 1, the bias is that the lineage of the population can come only from the splitting assembly by definition.

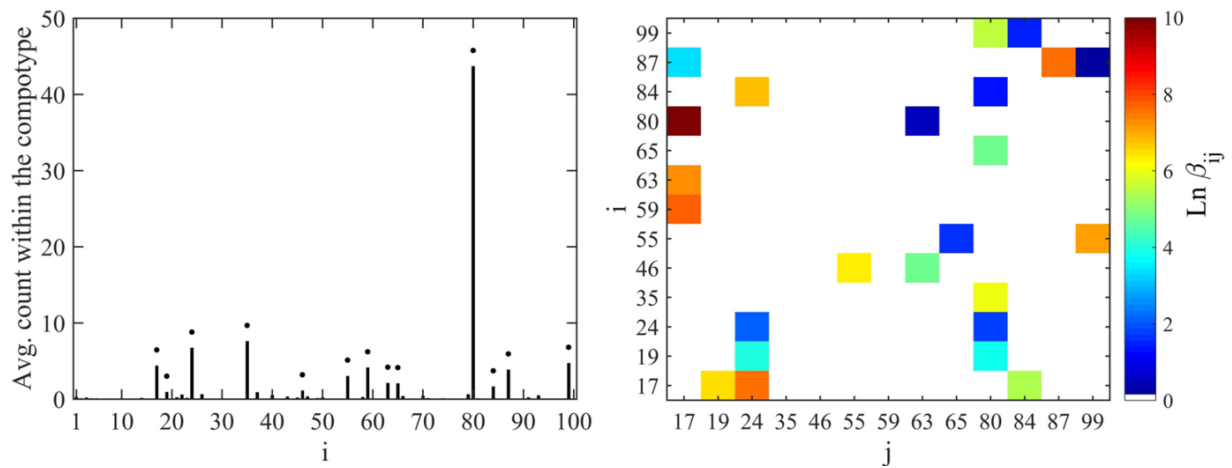

**Figure 3: A compotype that emerged under the β studied.** Left : Bar plot showing the average count of a given molecular-type (i=1..$N_G$) in the compotype (i.e. the compotype's center of mass). Black dots mark the top molecule-types that are found to consist of its internal repertoire (i.e. molecule-types with an average copy number larger than 1.0 (Markovitch and Lancet 2014)). Right: The sub-matrix of β for the internal repertoire.

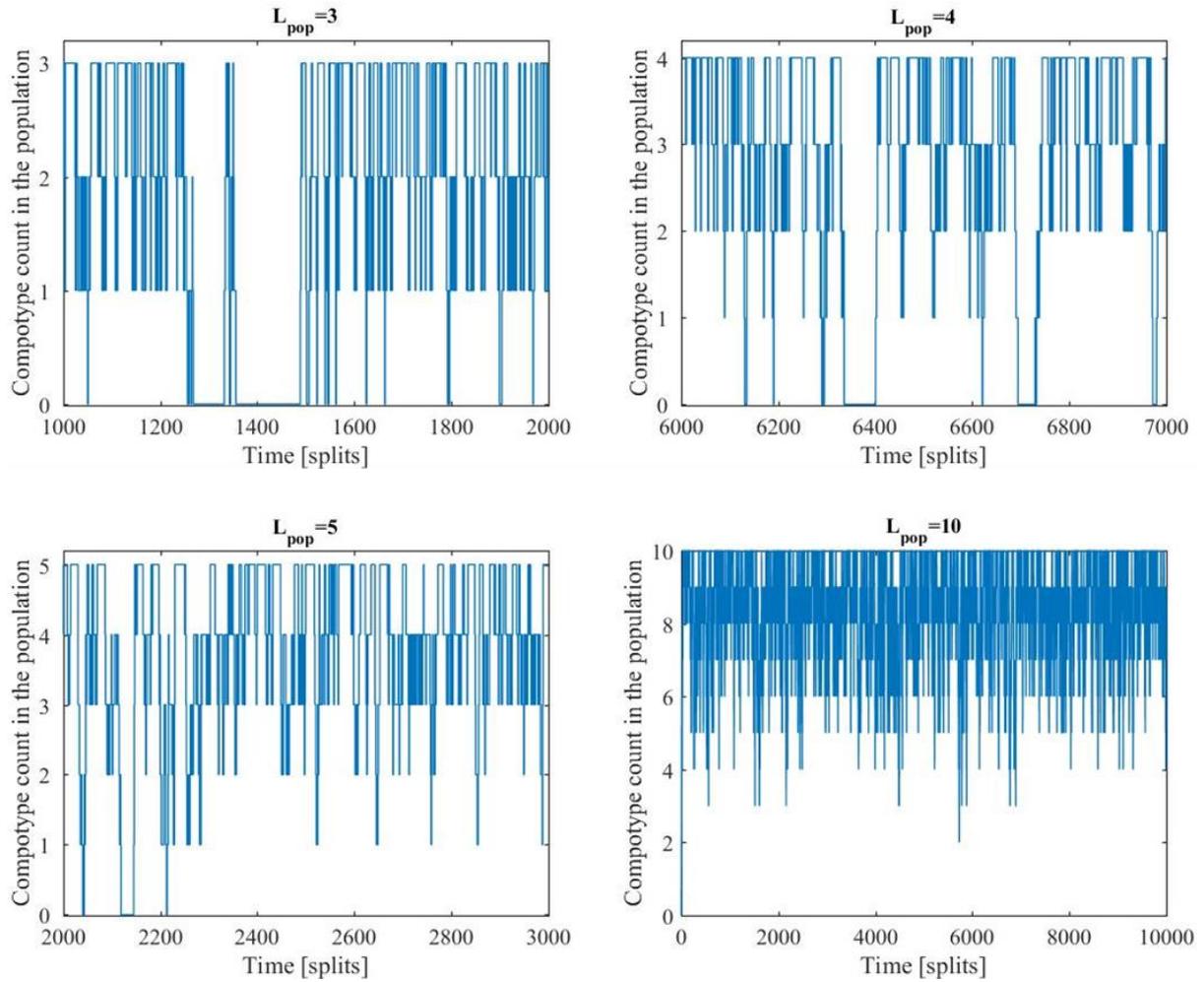

**Figure 4: Examples of GARD population dynamics with different population sizes (number of assemblies in the population).** Time is given in number of split events occurred in the system. At each time point, the content of each assembly in the system is compared with the compotype's center of mass to determine the number of assemblies that are part of the compotype's sub-population. For each population size, different time frames are picked in order to show significant dynamics in the simulation. All simulations performed with exactly the same parameter values and under the same β. Averages are presented in Figure 5. Full dataset is given in (Markovitch, Witkowski et al. 2018).

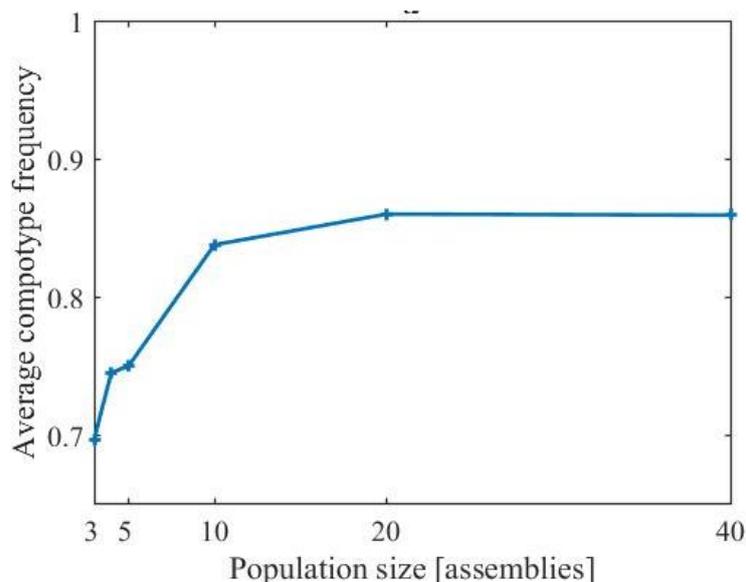

**Figure 5: Average frequency of the compotype in populations of increasing size.** A compotype's frequency is its count within the population divided by the population size. Data is averaged over the entire trajectory of each population.

### 6.2. Multilevel Price equation in GARD

This section analyses GARD's population dynamics (see Figure 4) from the point of view of the multilevel Price equation in order to understand how group selection is governing the system's dynamics, with the molecules being the equivalent of the individuals and the assemblies of the groups. As discussed above (section 4) the reactant-product chemistry analogy of the parent-progeny biology relationship is employed here.

In order to calculate the values of $w_{ij}$ in Equation 2, the parent-progeny relationships between all the assembly-dwelling molecules in the system is tracked. The specific formulation of the assembly growth process used here (Equation 3) enables determining, for each single copy of a molecule that joined an assembly, which specific copy of a molecule within the assembly recruited it. The recruiter takes the role of the parent and the molecule that joins takes the role of the progeny. Additionally, we track the time trajectory of each molecule belonging to the assemblies before and after the split (Figure 2).

We choose $z_{ij}=1$ for any molecule at any time if its molecule-type is part of the internal molecular repertoire of the compotype (Figure 3). The reason for this choice of z is that it has the properties: (1) $<z_i>=1$, marks that this assembly has the same repertoire (molecule-types) of the compotype[1]; (2) positive $cov_i(<w_i>, <z_i^A>)$ means that when an assembly acquires the classification of the compotype it reproduces faster on average (i.e. selection for groups); (3) positive $<cov_j(w_{ij}, z_{ij}^A)>_i$ means that, on average, inside each group there are molecules that recruit more molecules (i.e. selection for individuals).

For the above-described choice of $z_{ij}$, Figure 6 shows the average Price terms, calculated over all splits occurring during a whole simulation, as described in Section.

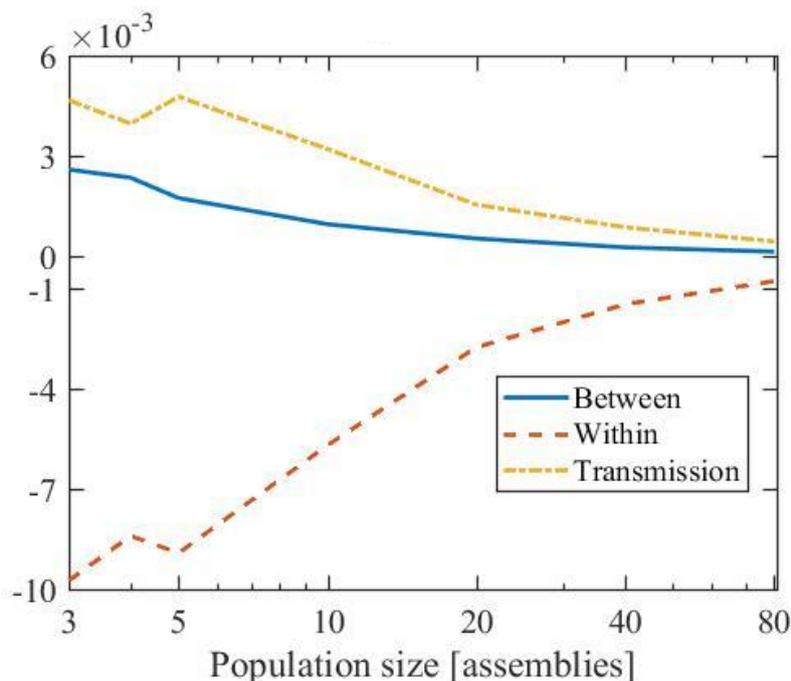

**Figure 6: The three terms in the Price equation (Equation 2) with respect to the compotype: Between, Within and Transmission.** For each population size, the Price equation is calculated after each split event with respect to the next split event. This figure shows the averages over all splits occurring during the whole simulation for each term.

## 7. Summary

We have demonstrated how evolutionary biology tools (the Price equation in this case) could be applied in a, complex, chemical system. Cooperation between the individual molecules, captured in the β matrix/network, leads to group-selection being favored.

GARD model was picked because it is a relatively simple model that give rise to complex behavior (i.e. replication and evolution), and has defined levels and generations. We effectively chose the form of $z_{ij}$ given that we have had prior knowledge on the emergence of compotype species, yet future works could extend this analysis by choosing different z. One such possibility, which potentially could be applied to other systems aswell, could be connected to entropy (given the intimate relation between information and entropy) (Mirmomeni, Punch et al. 2014).

---

[1] Note that a compotype is the specific ratio of the copies of the different molecule-types (i.e. ts composition), hence $<z_i>=1$ does not automatically mean that this assembly is the compotype.

Arguing that it is possible to apply biology tools on chemical systems, together with the present example help portray a route to further deepen the connections between chemistry and biology, as much as such connections exist.

For the origins of life, this supports a view of a continuous transition from the prebiotic chemistry towards present-day biology. The picture we imagine is that selection on ensembles of molecules gives rise to cooperative processes between molecules; this provides a starting point from which the complexity of biology can develop.

## 8. Acknowledgements

This project was supported by the Earth-Life Science Institute Origins Network, which is supported by a grant from the John Templeton Foundation. We thank Takashi Ikegami and Lana Sinapayen for discussions. OM acknowledges support from the Dutch Origins Center. We thank Lucy Kwok for help with drawing of Figure 2.

## 9. Appendix

9.1. The terms in Equation 2 can be explicitly written as (Kentzoglanakis, Brown et al. 2016):

$$\langle w_i \rangle = \frac{\sum_{ij} w_{ij}}{n_i^A}$$

Equation 4

$$\langle z_i^A \rangle = \frac{\sum_{ij} z_i^A}{n_i^A}$$

Equation 5

$$cov_i(\langle w_i \rangle, \langle z_i^A \rangle) = \frac{\sum_i n_i^A \langle w_i \rangle \langle z_i^A \rangle}{\sum_i n_i^A} - \left(\frac{\sum_i n_i^A \langle w_i \rangle}{\sum_i n_i^A}\right)\left(\frac{\sum_i n_i^A \langle z_i^A \rangle}{\sum_i n_i^A}\right)$$

Equation 6

$$\langle cov_j(w_{ij}, z_{ij}^A) \rangle_i = \frac{1}{\sum_i n_i^A} \sum_i n_i^A \left(\frac{\sum_j w_{ij} z_{ij}^A}{n_i^A} - \left(\frac{\sum_j w_{ij}}{n_i^A}\right)\left(\frac{\sum_j z_{ij}^A}{n_i^A}\right)\right)$$

Equation 7

Where $n_i^A$ is the size of ancestor group i.

9.2. We provide the full dataset of the simulations presented in this work ((Markovitch, Witkowski et al. 2018) https://zenodo.org/record/1172467). The format of the dataset is as follows:

I. Compotypes.txt; The average molecular count of each molecule-type (i=1..$N_G$) of the compotype studied.

For each system, its state at each time point (i.e. when an assembly reached its maximal size ($N_{max}$=100)):

II. Pop-X-NG.txt (X=number of assemblies in the population); The molecule-type index (i) of each molecule in each assembly. A value of '0' means no molecule. In each line, integers 1..100 hold the indices of the first assembly, 101..200 the indices of the second assembly, etc'.

III. Pop-X-Parents.txt (X=number of assemblies in the population); For each molecule, the index of its parent molecule.

IV. pr(m, t) is the parent index of molecule m at time t:
```
If m did not change during the split, pr=m.
If m is directly the result of the previous split, pr =
index of m before the previous split.
If m joined after a split, pr = index of its parent.
Otherwise, pr=0.
Note that it is needed first the verify if there is a
molecule ('Pop-X-NG.txt') before tracing its parent.
```

V. Beta.txt; The assumed chemical network (generated at random) used for all the simulations performed. For reference β(i=3, j=2)=1.157.